\RequirePackage{braket}

\documentclass[aps,prb,twocolumn,superscriptaddress,floatfix]{revtex4-2}

\usepackage{amsmath, amssymb, graphicx}
\usepackage[colorlinks=true,citecolor=blue,urlcolor=blue,linkcolor=blue]{hyperref}
\usepackage{xcolor}
\usepackage{braket}
\usepackage[utf8]{inputenc}
\usepackage{bm}

\usepackage{booktabs}

\newcommand{\Tr}{\operatorname{Tr}}

\begin{document}
\title{Mixed-valence fractional Chern insulators and fractionalized Fermi liquid in a topologically trivial band}
\author{Boran Zhou}
\author{Ya-Hui Zhang}
\affiliation{William H. Miller III Department of Physics and Astronomy, Johns Hopkins University, Baltimore, Maryland 21218, USA}
\begin{abstract}
Fractional Chern insulators (FCIs) are topologically ordered states characterized by a fractionally quantized Hall conductivity and a finite bulk gap, conventionally realized in partially filled narrow Chern bands. In this work, we challenge this convention by considering a topologically trivial band at filling $\nu=\frac{1}{3}$. In the flat-band limit, we obtain an exact FCI ground state with $\sigma_{xy}=-\frac{2e^2}{3h}$. By tuning the dispersion of the parent band, we identify a topological phase transition to another FCI with $\sigma_{xy}=\frac{e^2}{3h}$ using exact diagonalization (ED) and density matrix renormalization group (DMRG). At the transition point, an emergent spinless Dirac fermion coexists with a topologically ordered background. Through spectral function calculations, we show that the Dirac fermion is formed by the physical electron $c$ and an emergent dark fermion $\psi$. Doping away from this critical point by either electron or hole  gives rise to a fractionalized Fermi liquid (FL$^*$) phase with a small Fermi surface.
\end{abstract}
\maketitle

\emph{Introduction.---} The fractional quantum Hall (FQH) effect arises from interacting electrons in partially filled Landau levels, featuring a fractionally quantized Hall conductivity and intrinsic topological order~\cite{Tsui1982,Laughlin1983,Wen1990,WenNiu1990}. The FQH effect can also emerge in the absence of an external magnetic field, known as the fractional quantum anomalous Hall (FQAH) effect, and the corresponding states in lattice systems are dubbed fractional Chern insulators (FCIs)~\cite{Tang2011,Sun2011,Neupert2011,Sheng2011,Regnault2011,Parameswaran2013,Bergholtz2013,Abouelkomsan2020,Ledwith2020,Repellin2020,Li2021,Crepel2023}. Recent experimental observations of FQAH effects in moir\'e materials~\cite{Cai2023,Zeng2023,Park2023,Xu2023,Lu2024} have stimulated great interest in understanding the conditions for stabilizing FCIs. Conventionally, FCIs are expected to arise in partially filled Chern bands with quantum geometry resembling that of a Landau level. Several criteria for identifying bands favorable for FCIs have been proposed, including uniform Berry curvature~\cite{Parameswaran2012,Roy2014}, the trace condition~\cite{Roy2014,Claassen2015,Jackson2015,Mera2021,Wang2021,Wang2023}, and vortexability~\cite{Ledwith2023}. However, it remains unclear what the necessary conditions for realizing FCIs are~\cite{Simon2015}. In particular, recent theoretical studies have demonstrated that FCIs can even emerge in topologically trivial bands with highly nonuniform Berry curvature distributions~\cite{Lin2026,Liu2026}. It is then natural to ask whether an FCI in a $C=0$ band can exhibit properties with no counterpart in a Landau level.

On the other hand, while Mott physics is conventionally discussed in trivial bands, its realization in topological bands has recently attracted considerable interest, particularly in twisted bilayer graphene (TBG)~\cite{Song2022,Shi2022,Ledwith2025b,Ledwith2025,Zhao2025a,Zhao2025b,Zhou2026a,Vituri2026,Hu2026,Wei2026,Nosov2026,Zhou2026b,Zhang2026a,Zhang2026b}. A physical picture from these studies is a mixed-valence structure, where there are dilute doublon and holon excitations around $\mathbf{k}=0$~\cite{Zhao2025b,Zhou2026b,Zhang2026a}. In the presence of particle-hole (PH) symmetry, a quantum Mott semimetal can be realized, featuring a spinful Dirac fermion coexisting with quantum fluctuating local moments~\cite{Zhou2026b}. Breaking PH symmetry can lead to distinct symmetric topological Mott insulators (STMIs)~\cite{Zhou2026a}. These developments motivate us to explore similar mixed-valence physics in FQH settings. In particular, we ask whether a Dirac fermion can coexist with a topologically ordered FQH background, and whether its mass term can give rise to distinct FCIs within a single trivial band.

To address this question, we consider a spinless topologically trivial flat band, whose wavefunction is dominated by a lowest Landau level (LLL) orbital $f$ over most of the Brillouin zone, but by a trivial orbital $c_2$ near $\mathbf{k}=0$. This $C=0$ band can be regularized by a three-band model, similar to the topological heavy fermion model for the $|C|=1$ band of TBG~\cite{Song2022,Shi2022}. For simplicity, we consider the Haldane pseudopotential interaction $V_1$~\cite{Haldane1983,Trugman1985} in the $f$ orbital and project it to the active band. At filling $\nu=\frac{1}{3}$, a natural starting point is the $\frac{1}{3}$ Laughlin state~\cite{Laughlin1983} in the $f$ orbital. However, the orbital structure of the flat band allows for charge transfers between the Laughlin background and the $c_2$ orbital near $\mathbf{k}=0$. Transferring an electron from the Laughlin state to the $c_2$ orbital leaves behind three Laughlin quasiholes. The resulting state can thus be understood in terms of exciton pairing between a $c_2$ electron and a bound state of three Laughlin quasiholes. We obtain an exact ground state with Hall conductivity $\sigma_{xy}=-\frac{2e^2}{3h}$, originating from the $\frac{e^2}{3h}$ contribution of the Laughlin background and an additional $-\frac{e^2}{h}$ contribution from the topological structure of the exciton pairing.

Furthermore, we introduce a potential $\Delta$ on the $c_2$ orbital and project it to the active band, which generates a dispersion. Increasing $\Delta$ makes it energetically costly to create the exciton. When $\Delta>\Delta_c$, the system enters another FCI with $\sigma_{xy}=\frac{e^2}{3h}$, which can be adiabatically connected to the $\frac{1}{3}$ Laughlin state in the $f$ orbital. At the transition point, a Dirac cone centered at $\mathbf{k}=0$ emerges on top of the Laughlin background. Remarkably, although we start from a single spinless band, the two components of the emergent Dirac fermion are the physical electron $c$ and a dark fermion $\psi$, which creates a bound state of three quasiholes in the Laughlin background. The exciton is therefore created by $c^\dagger\psi$, and the weak-to-strong pairing transition of the excitons, analogous to the Bardeen--Cooper--Schrieffer (BCS) to Bose--Einstein-condensation (BEC) transition of a $p+\mathrm{i}p$ superconductor~\cite{ReadGreen2000}, is equivalently a sign change of the mass of the $(c,\psi)$ Dirac fermion, which changes $\sigma_{xy}$ by $\frac{e^2}{h}$. We provide an effective field theory of this Dirac fermion near the transition. One remarkable prediction follows from this two-component picture: for the two FCIs close to the transition, the cheapest charged excitation upon doping is the charge-$e$ Dirac fermion formed by $c$ and $\psi$, rather than deconfined anyons. As a consequence, upon electron or hole doping, we find a fractionalized Fermi liquid (FL$^*$)~\cite{Senthil2003,Senthil2004,Paramekanti2004} with a small Fermi surface whose volume counts only the doped carriers. To the best of our knowledge, this is the first clear microscopic realization of an FL$^*$ phase in a one-band model. Interestingly, in this FL$^*$ phase, the small pocket is formed together by the electron $c$ and the dark fermion $\psi$, similar to the ancilla theory of FL$^*$ for underdoped cuprates~\cite{Zhang2020a,Zhang2020b}.

We finally generalize our theory to a $C=2$ band, where we obtain three distinct FCIs at filling $\nu=\frac{1}{3}$ with Hall conductivities $\sigma_{xy}=\frac{e^2}{3h}$, $\frac{2e^2}{3h}$ and $\frac{4e^2}{3h}$ by tuning $\Delta$ and $s^2$, the fraction of the Brillouin zone  where the Berry curvature is concentrated. FCIs in higher Chern bands are commonly understood as color-entangled generalizations of Halperin states~\cite{Halperin1983,Barkeshli2012,Liu2012,Wang2012,Wu2013,Sterdyniak2013,Dong2023}, and conventionally only the Halperin-112 state with $\sigma_{xy}=\frac{2e^2}{3h}$ is expected at $\nu=\frac{1}{3}$ of a $C=2$ band~\cite{Niu2025,SongChang2025}. We find this state when $s^2$ is large. In contrast, the $\sigma_{xy}=\frac{e^2}{3h}$ state corresponds to the Laughlin state in the $C=1$ $f$ orbital, similar to a Laughlin phase recently found in a nonideal $C=2$ band~\cite{SongChang2025}, while the $\sigma_{xy}=\frac{4e^2}{3h}$ state arises from the topological exciton pairing on top of the Laughlin background. These results provide unconventional FCIs at $\nu=\frac{1}{3}$ filling of a $C=2$ band and further demonstrate the applicability of our mixed-valence picture to higher Chern bands.

Our results reveal a common feature of realizing strongly correlated states in narrow bands: a mixed-valence structure allows a physical electron and a dark fermion to form a Dirac cone within a single isolated band, coexisting with a Mott insulator or an FCI. In the present case, the background itself hosts intrinsic topological order with fractionalized excitations. More generally, our results suggest that mixed-valence physics offers a general route to realizing correlated states that are not dictated by the topology of the parent band.
\begin{figure}[t]
\includegraphics[width=0.95\linewidth]{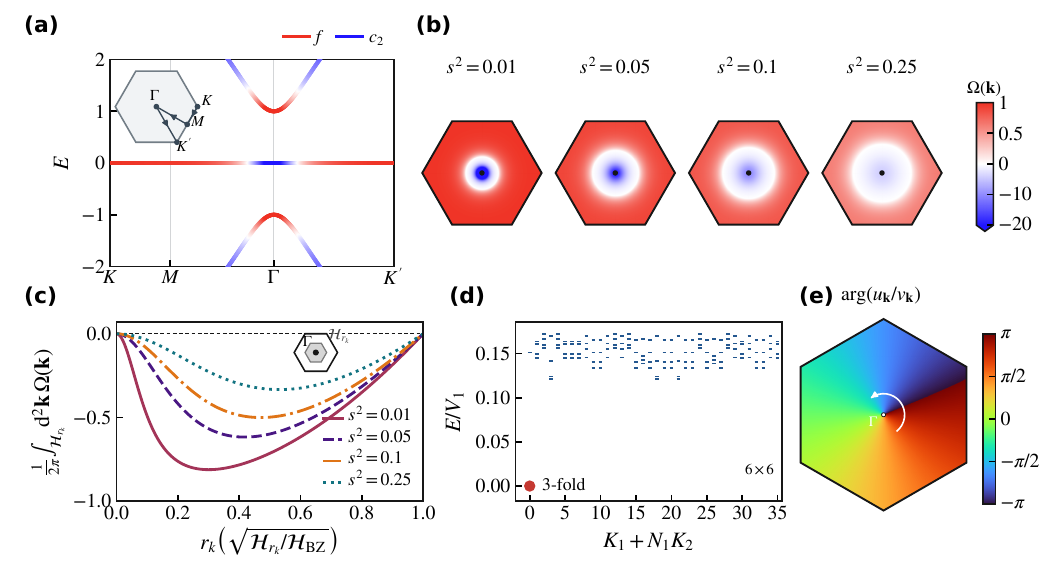}
\caption{\label{fig:illustration} Band structure, Berry curvature and manybody spectrum at $\Delta=0,\gamma=1$. (a) Band dispersion linecut. Red and blue indicate $f$ and $c_2$ components, respectively. (b) Berry curvature distribution for the indicated $s^2$, shown on a common color scale. (c) Berry curvature integrated over the shaded $\Gamma$-centered hexagon and divided by $2\pi$, versus $r_k$. (d) Low energy spectrum of the projected $V_1$ interaction for $N=12$ on a $6\times6$ torus at $s^2=0.05$. (e) $\mathrm{arg}(u_\mathbf{k}/v_\mathbf{k})$ shows an $p-\mathrm{i}p$ structure of exciton $c^\dagger_{2;\mathbf{k}}f_\mathbf{k}$ pairing.}
\end{figure}

\emph{Model.---}We start from a three-band spinless model, with orbitals $c_1$, $c_2$ and $f$. The $c_1$ and $f$ orbitals form two identical lowest Landau levels (LLLs), each with Chern number $+1$, while $c_2$ is a trivial orbital defined on a triangular lattice. The non-interacting Hamiltonian can be written as:
\begin{equation}
 H_0=\sum_{\bm k}\bm\Psi_{\bm k}^\dagger
 \begin{pmatrix}
 0&V_{\bm k}&\gamma\\
 V_{\bm k}^*&\Delta&0\\
 \gamma&0&0
 \end{pmatrix}
 \bm\Psi_{\bm k},
 \label{eq:threeorbital}
\end{equation}
where $\Psi_{\mathbf{k}}=(c_{1;\mathbf{k}},c_{2;\mathbf{k}},f_{\mathbf{k}})^T$ denotes the three orbital structure. $\gamma$ denotes a constant coupling between $c_1$ and $f$ orbitals. $V_\mathbf{k}$ is the hybirdization between $c_1$ and $c_2$, contains a node due to the Chern number difference, we choose the node to be at $\mathbf{k}=0$ and $V_\mathbf{k}\rightarrow0$ as $\mathbf{k}\rightarrow0$. We take $|V_\mathbf{k}|^2=\frac{\sqrt{3}\gamma^2}{6\pi s^2}(3-\sum_{i}\cos(\mathbf{k}\cdot\mathbf{a_i}))$. $\mathbf{a_i}$ represents the nearest neighbor bonds and $s$ is an additional parameter to control the concentration of $c_2$ orbital. We add an additional potential term on $c_2$ orbital, denoted by $\Delta$. 

We first focus on the case $\Delta=0$, the free band structure can be exactly solved. The band structure is shown in Fig.~\ref{fig:illustration}(a). The middle band is exactly flat and the wavefunction can be written as
\begin{equation}\label{eq:band}
    \lvert c_\mathbf{k}\rangle=\frac{\gamma\lvert c_{2;\mathbf{k}}\rangle-V_\mathbf{k}\lvert f_\mathbf{k}\rangle}{E_\mathbf{k}}=u_\mathbf{k}\lvert c_{2;\mathbf{k}}\rangle+v_\mathbf{k}\lvert f_\mathbf{k}\rangle,
\end{equation}
where $E_\mathbf{k}=\sqrt{\gamma^2+|V_\mathbf{k}|^2}$, and $\lvert c_\mathbf{k}\rangle=c^\dagger_\mathbf{k}\lvert0\rangle$, $\lvert c_{2;\mathbf{k}}\rangle=c^\dagger_{2;\mathbf{k}}\lvert0\rangle$ and $\lvert f_\mathbf{k}\rangle=f^\dagger_\mathbf{k}\lvert0\rangle$ are single-particle states. At $\mathbf{k}=0$, the orbital is purely $\lvert c_{2;\mathbf{k}}\rangle$. For small $s^2$, the Berry curvature is nearly uniform and the band is close to LLL over most of the Brillouin zone except a small region with fraction $s^2$ around $\mathbf k=0$, as shown in Fig.~\ref{fig:illustration}(b). The integrated negative curvature near $\mathbf{k}=0$ cancels the positive contribution from the rest of the Brillouin zone, yielding a vanishing band Chern number. We treat this band as the active band. The remote band gap is $\gamma$ while $\Delta=0$. We assume the interaction is only in $f$ orbitals, written as:
\begin{equation}\label{eq:Hintf}
    H_\mathrm{int}=\sum_{\mathbf{k_i},\mathbf{k_j},\mathbf{k_l},\mathbf{k_m}} W_{ijlm}f^\dagger_{\mathbf{k_i}}f^\dagger_{\mathbf{k_j}}f_\mathbf{k_l}f_\mathbf{k_m},
\end{equation}
where we take $W_{ijlm}$ to be the Haldane $V_1$ pseudopotential term, which can be written as:
\begin{equation}
\begin{split}
    W_{ijlm}=&\frac{2\pi l_B^2V_1}{A}\sum_\mathbf{q}(1-q^2l_B^2)e^{-\frac{q^2l_B^2}{2}}F_{\mathbf{k_i},\mathbf{k_m};\mathbf{q}}F_{\mathbf{k_j},\mathbf{k_l};-\mathbf{q}},\\
    F_{\mathbf{k_1},\mathbf{k_2};\mathbf{q}}=&\sum_\mathbf{G}\eta_\mathbf{G}\delta_{\mathbf{q},\mathbf{k_1}-\mathbf{k_2}-\mathbf{G}}e^{\frac{\mathrm{i}l_B^2}{2}((\mathbf{k_1}+\mathbf{k_2})\times\mathbf{G}+\mathbf{k_1}\times\mathbf{k_2})},
\end{split}
\end{equation}
where $A=2\pi l_B^2 N_{\mathrm{uc}}$ is the system area, $l_B$ is the magnetic length. We set $l_B=1$. $\mathbf{G}=n_1\mathbf{b_1}+n_2\mathbf{b_2}$, $\mathbf{b}_1=\frac{2\pi^\frac{1}{2}}{3^\frac{1}{4}}(\frac{\sqrt{3}}{2},-\frac{1}{2})^T, \mathbf{b}_2=\frac{2\pi^\frac{1}{2}}{3^\frac{1}{4}}(0,-1)^T$. $\eta_\mathbf{G}=(-1)^{n_1+n_2+n_1n_2}$ is the sign factor.
We focus on the projective limit with $\gamma\gg V_1,\Delta$. We project the interaction and $c_2$ to active band, obtaining the effective model:
\begin{equation}\label{eq:Heff}
    \begin{split}
    H_\mathrm{eff}=&\sum_{\mathbf{k_i},\mathbf{k_j},\mathbf{k_l},\mathbf{k_m}} W_{ijlm}v^*_\mathbf{k_i}v^*_\mathbf{k_j}v_\mathbf{k_l}v_\mathbf{k_m}c^\dagger_{\mathbf{k_i}}c^\dagger_{\mathbf{k_j}}c_\mathbf{k_l}c_\mathbf{k_m}\\
    &+\sum_\mathbf{k}\Delta |u_\mathbf{k}|^2 c^\dagger_{\mathbf{k}}c_\mathbf{k}.
    \end{split}
\end{equation}
In this model $\mathbf{k}=0$ is completely decoupled with other part, therefore we can treat $n_{\mathbf{k}=0}$ as a good quantum number and perform the exact diagonalization (ED) calculation.

\emph{Exact solution.---} We construct exact ground states of the effective Hamiltonian in Eq.~\ref{eq:Heff} in the flat band limit $\Delta=0$. First we let $H_{V_1}=\sum_\lambda A_\lambda^\dagger A_\lambda$ denote the pure $V_1$ parent Hamiltonian obtained from $H_{\mathrm{int}}$ by replacing the $f$ operators with the active-band operators $c$. We denote a Laughlin zero mode of $H_{V_1}^{(c)}$ by $|\phi_0\rangle$, and its counterpart with the same Fock coefficients in the $f$ basis by $|\phi_0\rangle_f$.

Then Eq.~\ref{eq:Heff} at $\Delta=0$ can be written as $\sum_\lambda \widetilde A_\lambda^\dagger\widetilde A_\lambda$, where $\widetilde A_\lambda$ is obtained from $A_\lambda$ by replacing each $c_{\mathbf k}$ with $v_{\mathbf k}c_{\mathbf k}$. For nonzero $v_{\mathbf k}$, we have $\widetilde A_\lambda=D^{-1}A_\lambda D$, with $D=\prod_{\mathbf k}v_{\mathbf k}^{n_{\mathbf k}}$ and $n_{\mathbf k}=c_{\mathbf k}^\dagger c_{\mathbf k}$. Since $A_\lambda|\phi_0\rangle=0$, we can have $\tilde{A}_\lambda D^{-1}|\phi_0\rangle=0$. Therefore a zero energy ground state can be written as
\begin{equation}\label{eq:Psi1}
    \lvert \Psi_1\rangle= \frac{1}{\mathcal{N}_1}D^{-1} \lvert \phi_0\rangle\propto\prod_\mathbf{k} \left(v_\mathbf{k}+u_\mathbf{k}c^\dagger_{2;\mathbf{k}}f_{\mathbf{k}}\right)\lvert \phi_0\rangle_f,
\end{equation}
where $\mathcal{N}_1$ is the normalization factor.
 This state can be viewed as an exciton condensate on top of a $\nu=\frac{1}{3}$ Laughlin state in the $f$ sector. The operator $c^\dagger_{2;\mathbf{k}}f_{\mathbf{k}}$ creates an exciton consisting of a $c_2$ electron bound to three Laughlin quasiholes. The exciton has an $p-\mathrm{i}p$ pairing as shown in Fig.~\ref{fig:illustration}(e). The exciton contributes total Berry curvature $-1$ to the parent Laughlin state, therefore giving a $\sigma_{xy}=-\frac{2e^2}{3h}$ final state. We provide the exact proof for the many body Chern number in the End Matter.
\begin{figure}[t]
\includegraphics[width=0.95\linewidth]{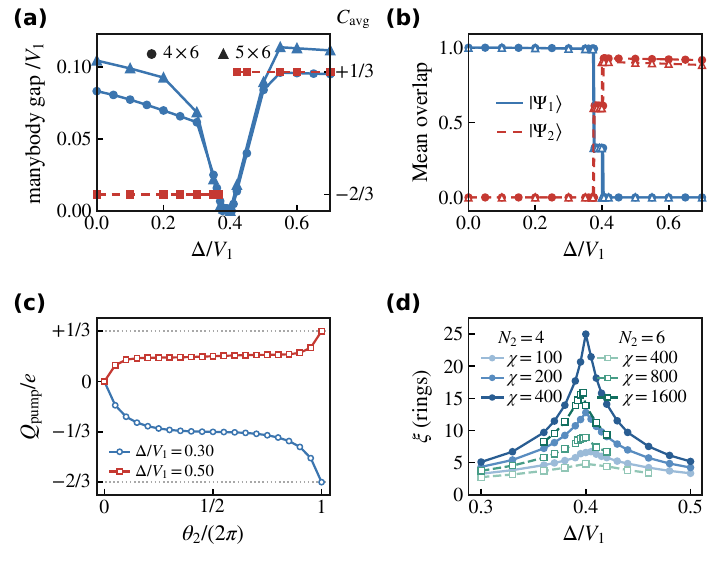}
\caption{\label{fig:transition} Parameters $s^2=0.05$, $\gamma=1$ and filling $\nu=\frac{1}{3}$. (a) Manybody gap and average Chern number $C_{\mathrm{avg}}$ of the lowest three states versus $\Delta/V_1$. (b) Average overlaps of $\lvert\Psi_1\rangle$ and $\lvert\Psi_2\rangle$ with the three degenrate ground states. (c) The insertion of $2\pi$ flux for parameters $\Delta/V_1=0.3,0.5$, corresponding to $\sigma_{xy}=-\frac{2e^2}{3h},\frac{e^2}{3h}$ respectively. (d) The neutral correlation length dependence on $\Delta/V_1$, which diverges at $\Delta=\Delta_c$. The results are obtained by iDMRG on infinite cylinders for (c)(d).  }
\end{figure}

\emph{Two FCIs.---} Tuning $\Delta$ can drive a transition to another FCI with $\sigma_{xy}=\frac{e^2}{3h}$, which can be adiabatically connected to the Laughlin state. We distinguish the two FCIs using the non-Abelian Chern number of the lowest three states under physical boundary twists $\bm\theta=(\theta_x,\theta_y)$~\cite{Niu1985}:
\begin{equation}\label{eq:manybodyChern}
 C_{\mathrm{avg}}=\frac{1}{6\pi}
 \int_0^{2\pi}\mathrm{d}\theta_x\int_0^{2\pi}\mathrm{d}\theta_y\,
 \Tr\mathcal F_{\theta_x\theta_y},
\end{equation}
$\mathcal F_{\theta_x\theta_y}$ is the non-Abelian Berry curvature of the manybody state, the details are in the End Matter. As shown in Fig.~\ref{fig:transition}(a), there is a transition point at $\Delta_c\approx0.396$, $C_\mathrm{avg}=-\frac{2}{3}$ for $\Delta<\Delta_c$ and $C_\mathrm{avg}=\frac{1}{3}$ for $\Delta>\Delta_c$. Although an exact solution is lack for the $\sigma_{xy}=\frac{e^2}{3h}$ FCI, we target a trial wavefunction:
\begin{equation}
    \lvert\Psi_2\rangle=\frac{1}{\mathcal{N}_2}D \lvert \phi_0\rangle,
\end{equation}
where $\mathcal{N}_2$ is the normalization factor. The state is the $\frac{1}{3}$ Laughlin state for $f$ projected to the active band. Compared to $\lvert\Psi_1\rangle$, we calculate the overlap with true ground state, as plot in Fig.~\ref{fig:transition}(b). The large overlap indicates the $C_\mathrm{avg}=\frac{1}{3}$ is approximately $\lvert\Psi_2\rangle$.

The change in $C_\mathrm{avg}$ by one suggests the existence of a Dirac cone at the transition. This transition can be understood as a weak
to strong pairing of the excitons. On the $\sigma_{xy}=-\frac{2e^2}{3h}$ side, the excitation of $c_2$ electron is cheap, therefore favoring a weak pairing of excitions. On the other hand, as $\Delta$ increases, creating an unbound extion pair becomes energetically costly, the system enters the strong pairing regime with $\sigma_{xy}=\frac{e^2}{3h}$. The transition can thus be viewed as a BCS-to-BEC transition. In our current Hamiltonian, the $c_{\mathbf k=0}$ mode is decoupled and its occupation $n_{\mathbf k=0}$ is a conserved quantum number. We have $n_{\mathbf k=0}=1$ for $\Delta<\Delta_c$ and $n_{\mathbf{k}=0}=0$ for $\Delta>\Delta_c$.

\begin{figure}[t]
\includegraphics[width=0.95\linewidth]{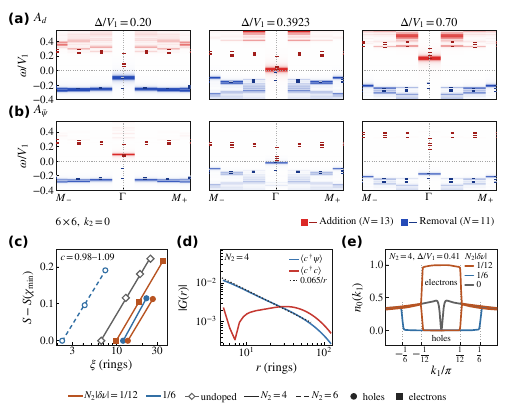}
\caption{\label{fig:Dirac} (a) Single electron spectral function $A_c$. (b) Trion spectral function $A_{\psi}$. (c) Entanglement entropy against the correlation length, undoped near the critical point at $\Delta/V_1=0.40$ ($N_2=4$) and doped at $\Delta/V_1=0.41$ ($N_2=4$) and $0.42$ ($N_2=6$). Fits to $S=\frac{c}{6}\ln\xi$ give $c=0.98$--$1.09$. The results indicate a Dirac cone at the critical point of $\nu=\frac{1}{3}$ and a small electron or hole pocket upon doping. (d) Electron–trion and electron–electron correlators at $\nu=\frac{1}{3}$ and the critical point $\Delta/V_1=0.405$. The dotted line shows $0.065/r$. (e) Momentum distribution $n_0(k_1)$ on the line $k_2=0$ in the doped FL$^*$ phase at filling $\nu=\frac{1}{3}+\delta\nu$, for $N_2=4$ and $\Delta/V_1=0.41$ (gray: undoped). The dotted lines mark $k_1=\pm\pi N_2|\delta\nu|$.  }
\end{figure}

\emph{DMRG results.---} To further characterize the two FCIs and the transition between them, we perform infinite density matrix renormalization group (iDMRG) calculations~\cite{Hauschild2018} on infinite cylinders that are periodic along $\mathbf a_2$, with circumferences of $N_2=4$ and $6$ unit cells. The momentum $k_2$ around the cylinder is then conserved, and we work in a hybrid basis that keeps $k_2$ but is localized along the cylinder: the Fourier transforms of the band states along the cylinder axis, which away from $\mathbf k=0$ are Landau-gauge orbitals of the LLL, as in iDMRG studies of quantum Hall and moir\'e systems~\cite{Zaletel2013,Motruk2016,Soejima2020} (details in the End Matter). By inserting a $2\pi$ flux through the cylinder, we find that the pumped charge is $-\frac{2e}{3}$ for $\Delta/V_1=0.3$ and $\frac{e}{3}$ for $\Delta/V_1=0.5$, as shown in Fig.~\ref{fig:transition}(c). This is consistent with the many-body Chern number from ED. Furthermore, as shown in Fig.~\ref{fig:transition}(d), the neutral correlation length $\xi$ diverges at the critical point, indicating a continuous transition.

\begin{figure}[!t]
\includegraphics[width=0.95\linewidth]{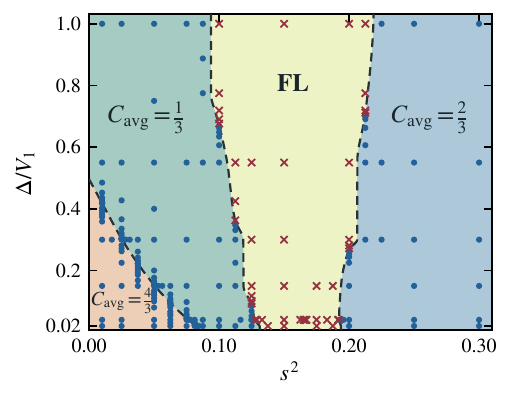}
\caption{\label{fig:higherChernband} Phase diagram of the $C=2$ band at $\nu=1/3$, with $V_1$ interactions in both $f$ and $c_2$ orbitals. The calculations are performed on a $5\times6$ torus by ED. }
\end{figure}

\emph{Dirac transition and dark fermion.---} As $\sigma_{xy}$ jumps by $\frac{e^2}{h}$ across the transition, we naturally expect a Dirac fermion at the critical point. However, we restrict to a single band and there is only one single-electron operator $c(\mathbf k)$, while a Dirac fermion needs two components.  The second component turns out to be emergent and is from a composite fermion. Although $f_\mathbf{k}$ component vanishes as $\mathbf{k}\rightarrow0$, the bound state of three Laughlin quasiholes can still be created by a trion operator in the active band:
\begin{equation}\label{eq:psi}
    \psi_\mathbf{k_i}=\sum_{\mathbf{k_j,k_l,k_m}\ne\mathbf{k_i}}\frac{W_{ijlm}v^*_{\mathbf{k_j}}v_{\mathbf{k_l}}v_{\mathbf{k_m}}}{(N-1)(W_{ijij}-W_{ijji})}c^\dagger_{\mathbf{k_j}}c_{\mathbf{k_l}}c_{\mathbf{k_m}},
\end{equation}
where $N$ is the number of particles. The details of the identification between trion operator and three bounded quasiholes are discussed in the End Matter. Since $\psi_\mathbf{k}$ is constructed from the Laughlin background of $f$ electrons, while $c_\mathbf{k}$ is dominated from the $c_2$ orbital near $\mathbf{k}=0$, the two fermions become orthogonal as $\mathbf{k}\rightarrow0$. Therefore, $\psi_\mathbf{k}$ is a dark fermion that is invisible in the electron spectral function, as shown in Fig.~\ref{fig:Dirac}(a). Separately, the trion spectral function in Fig.~\ref{fig:Dirac}(b) confirms that $\psi_\mathbf{k}$ contributed to the Dirac cone structure. Near the transition point, we therefore can write down an effective theory as 
\begin{equation}
    \mathcal{L}_\mathrm{eff}=\bar{\chi}(\mathrm{i}\gamma^\mu D_\mu-m)\chi-\frac{3}{4\pi}a\mathrm{d}a+\frac{1}{2\pi}A\mathrm{d}a,
\end{equation}
where $D_\mu=\partial_\mu-\mathrm{i}A_\mu$, $\chi$ is the Dirac spinor $(c,\psi)^T$. $\gamma^\mu$ are the Dirac gamma matrices in 2D. $m\approx\Delta-\Delta_c$ is the effective mass.  Our iDMRG results support this two-component Dirac fermion. At the critical point, the entanglement entropy grows as $S=(c/6)\ln\xi$ with central charge $c=1$~\cite{Pollmann2009} [Fig.~\ref{fig:Dirac}(c)], as expected for a single Dirac fermion on the cylinder, where only the momentum line through $\mathbf k=0$ is gapless. Moreover, the equal-time correlator between the electron $c$ and the dark fermion $\psi$ of Eq.~\eqref{eq:psi} decays as $1/r$ [Fig.~\ref{fig:Dirac}(d)], the power law of a massless Dirac fermion in one dimension, while the correlator of the electron with itself shows no such power law. The physical electron and the dark fermion thus form the two components of the critical Dirac fermion.

\emph{FCIs in higher Chern bands.---} The mixed-valence construction also applies to bands with higher Chern numbers. By flipping the sign of the Berry curvature in the small region around $\mathbf{k}=0$ with area fraction $s^2$, and keeping the band dominated by the $f$ orbital away from $\mathbf{k}=0$, we obtain a $C=2$ band. At $\nu=\frac{1}{3}$ of a $C=2$ band, a Halperin-112 state with $\sigma_{xy}=\frac{2e^2}{3h}$ is usually expected~\cite{Niu2025,SongChang2025}, and a Laughlin state with $\sigma_{xy}=\frac{e^2}{3h}$ was recently found in a nonideal band~\cite{SongChang2025}. Fig.~\ref{fig:higherChernband} shows the phase diagram from ED, with $V_1$ interactions in both of the $f$ and $c_2$ orbitals. At small $s^2$ and small $\Delta$, the excitons again condense on top of the Laughlin state of $f$, now adding $\frac{e^2}{h}$ to its Hall conductivity, which gives $\sigma_{xy}=\frac{4e^2}{3h}$. Increasing $\Delta$ suppresses the excitons, and the system enters the Laughlin state with $\sigma_{xy}=\frac{e^2}{3h}$, as for the $C=0$ band. Increasing $s^2$ makes the Berry curvature distribution more uniform, and the ground state becomes the Halperin-112 state with $\sigma_{xy}=\frac{2e^2}{3h}$. At intermediate $s^2$ the two regimes are separated by a region without an FCI, marked FL in Fig.~\ref{fig:higherChernband}.

\emph{Discussion.---} In conclusion, we have constructed a model with a single $C=0$ band that realizes two FCIs, with $\sigma_{xy}=-\frac{2e^2}{3h}$ and $\frac{e^2}{3h}$, separated by a Dirac transition. The Dirac fermion at the transition is formed by the physical electron $c$ and the dark fermion $\psi$, and doping it yields an FL$^*$ with a small Fermi pocket. These results are an FQH analog of the mixed-valence Mott insulators~\cite{Zhao2025a,Zhao2025b,Zhou2026a} and the quantum Mott semimetal~\cite{Zhou2026b} in topologically nontrivial bands, and they suggest a unified picture of mixed-valence bands. One orbital dominates the band over most of the Brillouin zone, where the interaction builds a correlated background, while another, free orbital near $\mathbf k=0$ hosts dilute charge fluctuations. The electron near $\mathbf k=0$ and an emergent dark fermion built from excitations of the background form a Dirac fermion, and the sign of its mass, rather than the topology of the parent band, selects the correlated insulator: distinct mixed-valence Mott insulators~\cite{Zhou2026a} on a background of local moments, and the two FCIs on a Laughlin liquid here. Doping such an insulator is therefore a natural route to FL$^*$ phases in single-band models. The two backgrounds, however, differ in robustness. In a mixed-valence Mott insulator, the doped itinerant carriers can have a large Kondo coupling to the local moments, and Kondo screening can destroy the Mott phase in favor of a heavy Fermi liquid~\cite{Zhang2026a,Zhang2026b}. The FCI background, in contrast, is gapped and has no local moments to screen, so we expect the mixed-valence FCI and its doped FL$^*$ to be more robust. Since our $C=0$ band admits localized Wannier orbitals $c_i$, the physics discussed here can also be realized in a one-orbital lattice model with short-range off-site interactions, which we leave for future work.

\textit{Note added.}---Upon finishing this manuscript, we became aware of several related works on FCIs~\cite{Zhao2026,HerzogArbeitman2026,Wang2026,LiuPo2026}.

\textit{Acknowledgement.}---We thanks Zhao Liu and Jing-Yu Zhao for helpful discussions. We used Claude Opus 5.5 and GPT 6 Astra for numerical calculations and language editing. This work was supported by the National Science Foundation under Grant No.~DMR-2237031.

\bibliography{references}
\par\vspace{1\baselineskip}
\bigskip
\bigskip
\begin{center}
\textbf{\large End Matter}
\end{center}
\bigskip
\vspace{-0.2cm}

\emph{Exact proof of $C_\mathrm{avg}=-\frac{2}{3}$.---} The exact wavefunction of the $\sigma_{xy}=-\frac{2e^2}{3h}$ state is given in Eq.~\eqref{eq:Psi1}. We define
\begin{equation}
    T=\prod_\mathbf{k}t_\mathbf{k},\qquad t_\mathbf{k}=v_\mathbf{k}+u_\mathbf{k}c^\dagger_{2;\mathbf{k}}f_\mathbf{k},
\end{equation}
with $u_\mathbf{k}$ and $v_\mathbf{k}$ the amplitudes of the band state in Eq.~\eqref{eq:band}. The operator $t_\mathbf{k}$ turns an $f$ electron into a band electron, $t_\mathbf{k}f^\dagger_\mathbf{k}\lvert0\rangle=v_\mathbf{k}f^\dagger_\mathbf{k}\lvert0\rangle+u_\mathbf{k}c^\dagger_{2;\mathbf{k}}f_\mathbf{k}f^\dagger_\mathbf{k}\lvert0\rangle=(u_\mathbf{k}c^\dagger_{2;\mathbf{k}}+v_\mathbf{k}f^\dagger_\mathbf{k})\lvert0\rangle=\lvert c_\mathbf{k}\rangle$, where $u_\mathbf{k}c^\dagger_{2;\mathbf{k}}+v_\mathbf{k}f^\dagger_\mathbf{k}=c^\dagger_\mathbf{k}$ creates the band state of Eq.~\eqref{eq:band}; on an empty momentum $f_\mathbf{k}$ gives zero, so $t_\mathbf{k}$ multiplies it by $v_\mathbf{k}$. The same holds with other electrons present, since $t_\mathbf{k}$ acts only on momentum $\mathbf k$. $T$ therefore maps each $f$ Fock state onto the band Fock state with the same occupied momenta, weighted by $v_\mathbf{k}$ for every empty momentum, which is the second form of Eq.~\eqref{eq:Psi1}. With the three Laughlin states $\lvert\phi_{0;a}\rangle_f$ of the $f$ orbital, $a=1,2,3$, the three degenerate ground states on the torus are
\begin{equation}
    T\lvert\phi_{0;a}\rangle_f=\mathcal{N}_a\lvert\Psi_{1;a}\rangle,
\end{equation}
where $\mathcal{N}_a^2=\langle\phi_{0;a}|_fT^\dagger T|\phi_{0;a}\rangle_f>0$: even when an allowed momentum sits at $\Gamma$, where $v_\mathbf{k}=0$, the configurations of $\lvert\phi_{0;a}\rangle_f$ in which this momentum is occupied survive.

The twists $\bm\theta$ of Eq.~\eqref{eq:manybodyChern} shift the allowed momenta, so the orbital states, $u_\mathbf{k}$ and $v_\mathbf{k}$ all depend on $\bm\theta$. Only the trace of the non-Abelian Berry curvature enters, $\mathrm{Tr}\mathcal F=\partial_{\theta_x}\mathrm{Tr}\mathcal A_{\theta_y}-\partial_{\theta_y}\mathrm{Tr}\mathcal A_{\theta_x}$, with $(\mathcal A_{\theta_\mu})_{ab}=\mathrm{i}\langle\Psi_{1;a}|\partial_{\theta_\mu}|\Psi_{1;b}\rangle$. Since the three states stay normalized and linearly independent for every $s^2$, $C_\mathrm{avg}$ cannot change with $s^2$, and we evaluate it for $s^2\to0$. Then $|u_\mathbf{k}|\to0$ and $|v_\mathbf{k}|\to1$ except in a vanishing neighborhood of $\Gamma$. Whenever no allowed momentum lies in this neighborhood, $T$ reduces to the phase factor $e^{\mathrm{i}\Lambda}$ with $\Lambda(\bm\theta)=\sum_\mathbf{k}\arg v_\mathbf{k}$, so that $\lvert\Psi_{1;a}\rangle=e^{\mathrm{i}\Lambda}\lvert\phi_{0;a}\rangle_f$. Then
\begin{equation}\label{eq:gaugeLambda}
\begin{split}
    (\mathcal A_{\theta_\mu})_{ab}&=\mathrm{i}\langle\phi_{0;a}|_fe^{-\mathrm{i}\Lambda}\partial_{\theta_\mu}\big(e^{\mathrm{i}\Lambda}|\phi_{0;b}\rangle_f\big)\\
    &=(\mathcal A^{(0)}_{\theta_\mu})_{ab}-\delta_{ab}\,\partial_{\theta_\mu}\Lambda,
\end{split}
\end{equation}
where $\mathcal A^{(0)}$ is the Berry connection of the Laughlin states. Hence $\mathrm{Tr}\mathcal A_{\theta_\mu}=\mathrm{Tr}\mathcal A^{(0)}_{\theta_\mu}-3\partial_{\theta_\mu}\Lambda$, and $\mathrm{Tr}\mathcal F=\mathrm{Tr}\mathcal F^{(0)}-3(\partial_{\theta_x}\partial_{\theta_y}-\partial_{\theta_y}\partial_{\theta_x})\Lambda=\mathrm{Tr}\mathcal F^{(0)}$ wherever $\Lambda$ is smooth.

This holds for all twists except a small disk $D$ around the twist $\bm\theta_0$ at which an allowed momentum $\mathbf k_*$ crosses $\Gamma$. On an $N_1\times N_2$ torus the allowed momenta are $\mathbf k=(m_1+\frac{\theta_x}{2\pi})\frac{\mathbf{b_1}}{N_1}+(m_2+\frac{\theta_y}{2\pi})\frac{\mathbf{b_2}}{N_2}$ with integers $m_{1,2}$, so as $\bm\theta$ covers $[0,2\pi)^2$ each of them sweeps one cell of the momentum grid, and exactly one sweeps through $\Gamma$. Inside $D$ the states $T\lvert\phi_{0;a}\rangle_f/\mathcal N_a$ are smooth, since $T$ depends smoothly on $\bm\theta$ and $\mathcal N_a>0$ (at $\bm\theta_0$ the momentum at $\Gamma$ is occupied by the $c_2$ orbital). Stokes' theorem therefore gives
\begin{equation}\label{eq:stokesD}
\begin{split}
    \int_D\mathrm{d}^2\theta\,\mathrm{Tr}\mathcal F&=\oint_{\partial D}\mathrm{d}\theta_\mu\,\mathrm{Tr}\mathcal A_{\theta_\mu}\\
    &=\oint_{\partial D}\mathrm{d}\theta_\mu\,\mathrm{Tr}\mathcal A^{(0)}_{\theta_\mu}-3\oint_{\partial D}\mathrm{d}\theta_\mu\,\partial_{\theta_\mu}\Lambda\\
    &=\int_D\mathrm{d}^2\theta\,\mathrm{Tr}\mathcal F^{(0)}-6\pi w.
\end{split}
\end{equation}
The second line uses Eq.~\eqref{eq:gaugeLambda} on $\partial D$, where the three states again equal $e^{\mathrm{i}\Lambda}\lvert\phi_{0;a}\rangle_f$. The third uses Stokes' theorem for the Laughlin states, which are smooth in $D$, and $\oint_{\partial D}\mathrm{d}\theta_\mu\,\partial_{\theta_\mu}\Lambda=2\pi w$, which we now derive.

The winding number of $v_\mathbf{k}$ around $\Gamma$ is
\begin{equation}\label{eq:winding}
    w=\frac{1}{2\pi}\oint_{C_\Gamma}\mathrm{d}k_\mu\,\partial_{k_\mu}\arg v_\mathbf{k},
\end{equation}
where $C_\Gamma$ is a small counterclockwise circle around $\Gamma$. In $\Lambda=\sum_\mathbf{k}\arg v_\mathbf{k}$, every term except $\arg v_{\mathbf k_*}$ is smooth and single valued in $D$, because only $v_{\mathbf k_*}$ vanishes there, so these terms give no contribution to the loop integral. Near $\bm\theta_0$, $\mathbf k_*(\bm\theta)=[(\theta_x-\theta_{0,x})\mathbf{b_1}/N_1+(\theta_y-\theta_{0,y})\mathbf{b_2}/N_2]/2\pi$ is linear in $\bm\theta$ and preserves orientation, so as $\bm\theta$ runs once counterclockwise around $\partial D$, $\mathbf k_*$ runs once counterclockwise around $\Gamma$, and
\begin{equation}
\begin{split}
    \oint_{\partial D}\mathrm{d}\theta_\mu\,\partial_{\theta_\mu}\Lambda&=\oint_{\partial D}\mathrm{d}\theta_\mu\,\partial_{\theta_\mu}\arg v_{\mathbf k_*(\bm\theta)}\\
    &=\oint_{C_\Gamma}\mathrm{d}k_\mu\,\partial_{k_\mu}\arg v_\mathbf{k}=2\pi w.
\end{split}
\end{equation}
To evaluate $w$, note that $u_\mathbf{k}=\gamma/E_\mathbf{k}>0$, so $\arg(u_\mathbf{k}/v_\mathbf{k})=-\arg v_\mathbf{k}$. Its $p-\mathrm{i}p$ structure in Fig.~\ref{fig:illustration}(e) means that near $\Gamma$, $\arg v_\mathbf{k}=\varphi_\mathbf{k}+\mathrm{const}$, with $\varphi_\mathbf{k}$ the polar angle of $\mathbf k$; Eq.~\eqref{eq:winding} then gives $w=1$. This is the winding that cancels the Chern number of the LLL and makes the band trivial.

Adding the twists outside $D$, where $\mathrm{Tr}\mathcal F=\mathrm{Tr}\mathcal F^{(0)}$, to Eq.~\eqref{eq:stokesD}, we get $\int\mathrm{d}^2\theta\,\mathrm{Tr}\mathcal F=\int\mathrm{d}^2\theta\,\mathrm{Tr}\mathcal F^{(0)}-6\pi w$ over the whole torus of twists. With Eq.~\eqref{eq:manybodyChern} and $C_\mathrm{avg}=\frac{1}{3}$ for the Laughlin states,
\begin{equation}
    C_\mathrm{avg}=\frac{1}{3}-w=-\frac{2}{3}.
\end{equation}
The exciton condensate thus removes exactly the Chern number of the LLL, independent of the details of $V_\mathbf{k}$.

\emph{Origin of $\psi$.---} In the main text, $\psi_\mathbf{k}$ of Eq.~\eqref{eq:psi} creates a bound state of three quasiholes on the Laughlin state. Here we show this analytically. Since the Laughlin state $\lvert\phi_0\rangle_f$ is a zero-energy ground state of the $V_1$ interaction of Eq.~\eqref{eq:Hintf},
\begin{equation}
    \sum_{\mathbf{k_l},\mathbf{k_m}}W_{ijlm}f_{\mathbf{k_l}}f_\mathbf{k_m}\lvert\phi_0\rangle_f=0,
\end{equation}
and with Eq.~\eqref{eq:psi} one verifies that
\begin{equation}
\begin{split}
    \psi_\mathbf{k_i}|\phi_0\rangle_f=&-\frac{1}{N-1}\sum_{\mathbf{k_j}\ne\mathbf{k_i}}f^\dagger_{\mathbf{k_j}}f_\mathbf{k_i}f_\mathbf{k_j}\lvert\phi_0\rangle_f\\
    =&\frac{1}{N-1}\sum_{\mathbf{k_j}\ne\mathbf{k_i}}n_{\mathbf{k_j}}f_\mathbf{k_i}\lvert\phi_0\rangle_f=f_\mathbf{k_i}\lvert\phi_0\rangle_f.
\end{split}
\end{equation}
In the first line, the zero-mode condition replaces, for each $\mathbf{k_j}$, the terms of Eq.~\eqref{eq:psi} by minus those with $\mathbf{k_l}$ or $\mathbf{k_m}$ equal to $\mathbf{k_i}$, which momentum conservation reduces to $(W_{ijij}-W_{ijji})f_\mathbf{k_i}f_\mathbf{k_j}$; this cancels the denominator. In the second line, $n_\mathbf{k}=f^\dagger_\mathbf{k}f_\mathbf{k}$, and the sum counts the $N-1$ electrons left after $f_\mathbf{k_i}$ acts.
Therefore, although $\psi_\mathbf{k_i}$ contains no $f_\mathbf{k_i}$, it produces the same state as $f_\mathbf{k_i}$ acting on the Laughlin state, i.e., a bound state of three quasiholes. To see how, we decompose the Laughlin state as
\begin{equation}
    \lvert \phi_0\rangle_f=\sqrt{\nu} f^\dagger_\mathbf{k} \lvert N-1\rangle_{f_\mathbf{k}}+\sqrt{1-\nu}\lvert N\rangle_{f_\mathbf{k}},
\end{equation}
where $\lvert N-1\rangle_{f_\mathbf{k}}$ and $\lvert N\rangle_{f_\mathbf{k}}$ are normalized states with $N-1$ and $N$ electrons in the other $f$ orbitals and $f_\mathbf{k}$ empty. The operator $f_\mathbf{k}$ annihilates $\lvert N\rangle_{f_\mathbf{k}}$ and maps $f^\dagger_\mathbf{k}\lvert N-1\rangle_{f_\mathbf{k}}$ to $\lvert N-1\rangle_{f_\mathbf{k}}$. Conversely, $\psi_\mathbf{k}$ annihilates $f^\dagger_\mathbf{k}\lvert N-1\rangle_{f_\mathbf{k}}$ and maps $\lvert N\rangle_{f_\mathbf{k}}$ to $\lvert N-1\rangle_{f_\mathbf{k}}$.

\emph{DMRG method.---} We study Eq.~\eqref{eq:Heff} with the parameters of Fig.~\ref{fig:transition}. The $c_2$ orbitals sit on the sites $\mathbf r=n\mathbf a_1+m\mathbf a_2$ of a triangular lattice, with primitive vectors $\mathbf a_1=a(\frac{\sqrt3}{2},\frac12)$ and $\mathbf a_2=a(0,1)$, two of the nearest-neighbor vectors $\mathbf{a_i}$ in $V_\mathbf{k}$; the lattice constant $a=(4\pi/\sqrt3)^{1/2}$ makes each unit cell hold one flux quantum. The cylinder is periodic along $\mathbf a_2$ with $N_2$ unit cells around it, so $k_2=\mathbf k\cdot\mathbf a_2$ takes the values $(2\pi j+\theta_2)/N_2$, where $\theta_2$ is a twist of the boundary condition. Each value of $k_2$ cuts the Brillouin zone along a line $j$ on which $k_1=\mathbf k\cdot\mathbf a_1$ varies; at $\theta_2=0$ the line $j=0$ has $k_2=0$ and passes through $\Gamma$. We use hybrid Wannier orbitals $d_{j,n}$~\cite{Motruk2016,Soejima2020}, the Fourier transforms over $k_1$ of the band states $\lvert c_\mathbf{k}\rangle$ at fixed $k_2$, localized near the ring $n$ formed by the $N_2$ sites with the same $n$. For $k_2\neq0$ they are Landau-gauge orbitals of the LLL dressed by the nearest $c_2$ ring; at $k_2=0$ they are shared between two neighboring Landau orbitals and the $c_2$ ring between them.

We order the orbitals into a one-dimensional chain, $d_{0,n},\dots,d_{N_2-1,n}$ ring after ring. In this basis the projected $V_1$ interaction is a sum of squares of pair annihilators, which we write as a matrix product operator on this chain~\cite{Zaletel2013,Soejima2020}. At each cut between neighboring orbitals of the chain, the interaction terms acting on both sides of the cut are decomposed by singular values within each sector of particle number and momentum, and singular values below $10^{-3}V_1$ are dropped. This leaves a matrix product operator of bond dimension about $85$ for $N_2=4$ and $140$ for $N_2=6$. We run two-site iDMRG~\cite{Hauschild2018} with conserved particle number and $k_2$, starting from the root pattern $100100\ldots$ of the Laughlin state. $\chi$ denotes the bond dimension of the matrix product state. We record the entanglement entropy $S$, the correlation lengths $\xi$ of neutral and $\xi_e$ of charge-$e$ excitations in units of the ring spacing $\frac{\sqrt3}{2}a$, and the occupation of each line, $n_j(k_1)=\sum_r e^{-\mathrm{i}k_1r}\langle d^\dagger_{j,n}d_{j,n+r}\rangle$. For the Hall conductance we raise $\theta_2$ from $0$ to $2\pi$ and measure the charge pumped across a cut, counted by its position along the cylinder~\cite{Laughlin1981,Zaletel2014}. 

\end{document}